\documentstyle[12pt]{article}
\newcommand{\be}{\begin{equation}}
\newcommand{\ee}{\end{equation}}
\newcommand{\bea}{\begin{eqnarray}}
\newcommand{\eea}{\end{eqnarray}}
\newcommand{\ba}{\begin{array}}
\newcommand{\ea}{\end{array}}

\newcommand{\norsl}{\normalsize\sl}
\newcommand{\norsc}{\normalsize\sc}
\begin{document}


\title
{One-loop correction to the $\gamma W W$ vertex \\
in the  $e^- \gamma$ collider 
\thanks{Talk presented by Y.Yasui at INS Workshop 
``{\it Physics of $e^+ e^-$,$e^- \gamma$ and $\gamma \gamma$
collisions at linear accelerators}''INS,Tokyo,Japan}
}

\author{
\norsc    Jiro KODAIRA, Hiroshi TOCHIMURA 
and Yoshiaki YASUI\\
\norsl  Department of Physics, Hiroshima University\\
\norsl  Higashi-Hiroshima 724, JAPAN\\
~\\
\norsc Isamu WATANABE\\
\norsl  Department of Physics, 
         Ochanomizu University \\
\norsl      Bunkyo-ku Tokyo 112, JAPAN \\
}

\date{}
\maketitle

\begin{abstract}
{\normalsize  
We apply the pinch technique 
, which is a method to construct 
the gauge independent off-shell Green's functions, 
to the process $e^-\gamma \rightarrow W^-\nu$ 
to study the effects of radiative corrections to 
$WW\gamma$ three gauge boson vertex.
The one-loop contributions to the anomalous gauge  
boson couplings are estimated in the standard model. 
}
\end{abstract}

\begin{picture}(5,2)(-320,-430)
\put(3,-30){HUPD-9509}
\put(3,-45){OCHA-PP-57}
\put(3,-60){March, 1995}
\end{picture}

%
\setcounter{page}{1}
\baselineskip 24pt

\section{Introduction}

Measuring the self-couplings of the electroweak gauge-bosons 
is one of the important topics at the next linear colliders. 
A general approach to examine the $WWZ$ and $WW{\gamma}$
vertices  was discussed by K.Hagiwara {\sl et.al} in terms of the 
effective Lagrangian.\cite{HAGI} 
They studied the process 
$e^+e^-{\rightarrow}W^+W^-$ with generalized $ZW$ and 
$\gamma W$ couplings. 
The explicit form of the effective Lagrangian is 
given by,
\bea
{\cal L}_{WWV}&=&ig_V[g^V_1(W^+_{\mu\nu}W^{\mu}V^{\nu}
-W^+_{\mu}V_{\nu}W^{\mu\nu})\nonumber\\
&+&{\kappa_V}W^+_{\mu}W_{\nu}V^{\mu\nu}
    +{\lambda_V\over m_W^2}
      W^+_{\lambda\mu}W^\mu_{\nu}V^{\nu\lambda}
]+\cdots~~~~~~~~,
\eea
where 
$A_{\mu\nu}=\partial_{\mu}A_\nu-\partial_{\nu}A_\mu$, 
$V=\gamma$ or $Z$ and 
$\tilde{V}_{\mu\nu}={1\over2}\varepsilon
_{\mu\nu\rho\sigma}V^{\rho\sigma}$. 
In particular, 
$\kappa_\gamma$ and $\lambda_\gamma$ are related 
to the magnetic dipole moment $\mu_W$ 
and the electric quadrupole moment $Q_W$ as; 
\be
\mu_W~=~{e \over 2m_W}
 (1+\kappa_\gamma+\lambda_\gamma)
\ee
\be
Q_W~=~-{e \over m_W^2}
(\kappa_\gamma-\lambda_\gamma)~~~~.
\ee
In the context of the standard model, 
$\kappa_\gamma=1$, $\lambda_\gamma=0$ at the tree level . 
A deviation from the standard model prediction for 
these quantities may suggest a new physics. 
So the radiative corrections to these quantities must 
be estimated before discussing new physics. 
J.Papavassiliou and K.Philippides 
used the pinch technique to construct 
the one-loop gauge-invariant 
${\gamma}W^+W^-$ vertex in the standard model.\cite{EEWW} 
They applied this result to the process 
$e^+e^-{\rightarrow}W^+W^-$,  
and estimated the contribution of the radiative 
correction to the anomalous couplings in the parameters 
$\kappa_\gamma$ and $\lambda_\gamma$.
 
Now $e^-\gamma$ and $\gamma\gamma$ colliders are seriously 
considered as the interesting options to upgrade 
the $e^+e^-$ next linear colliders.\cite{GGCO} 
E.Yehudai investigated the effects of the anomalous couplings  
to the processes $e^-\gamma\rightarrow{\nu}W^-$ and 
$\gamma\gamma\rightarrow{W^+}W^-$ in 
the context of the effective Lagrangian.\cite{YEHU} 
In these papers, it has been pointed out that using the process 
$e^-\gamma\rightarrow{\nu}W^-$ 
has several advantages in measuring $W\gamma$ couplings.
 From refs.\cite{HAGI}\cite{YEHU}, we get the allowed regions
in the $\kappa_\gamma$-$\lambda_\gamma$ plane for the 
experimental measurements at the next linear colliders.

This talk contains two parts of topics.
First we review the S-matrix pinch 
technique and explain how to construct the 
gauge-invariant self-energy of the gauge-boson 
in the QCD case as a simplest example. 
Next we apply the pinch technique 
to the process $e^-\gamma\rightarrow{\nu}W^-$ 
to study the electroweak one-loop correction to the gauge 
independent $WW\gamma$ vertex.
We examine the behavior of the $WW\gamma$ vertex 
at the $e^-\gamma$ collider.

\section{Pinch technique}
Using the pinch technique we get 
the gauge-invariant off-shell Green's functions. 
In  this section we briefly review the S-matrix pinch 
technique.\cite{SMPT} 
We show the outline to construct the gauge-invariant 
two-point function of a gauge boson in the QCD case, 
as a simplest example.
The case of spontaneously symmetry broken 
theories was discussed in refs.\cite{SSB}.

Now we consider the two-body scattering of two test quarks.
The S-matrix element $T$ is gauge independent 
to any order in the perturbation theory.
We can decompose the S-matrix element 
$T$ in the form, 
\be
  T(t,s,m_1,m_2)
     ~=~A(t)~+~B(t,m_1,m_2)~+~C(t,s,m_1,m_2)~~~,
\label{smp1}
\ee
where $m_1$ and $m_2$ are masses of the test quarks, $t$ 
and $s$ are the Mandelstam variables, 
\be
    t~=~-(p'_1-p_1)^2~=~-q^2~~~~
         {\rm and}~~~s~=~(p_1+p_2)^2~~.
\nonumber
\ee
The functions $A$, $B$ and $C$ are evidently gauge
independent. From the different kinematical structure 
of these functions, we have possibility to define 
new Green's function consulting eq.(\ref{smp1}). 
For example, the gauge-independent function $A(t)$ can be 
considered as the new two-point function.(fig.1) 

The explicit calculation at the one-loop level 
is as follows. It is noted that 
the gauge dependence of the self-energy part fig.2(a) is 
canceled by the propagator-like contribution from 
fig.2(b) and fig.2(c) which we call ``pinch part''.
For simplicity, we choose the Feynman-t'Hooft gauge ($\xi=1$) 
in the following calculation. 
In this gauge, there are no pinch parts 
from the box diagrams fig.2(c).
The three gluon vertex $\Gamma_{\alpha\mu\beta}$ is given
by, (in the following, we omit the group factors.)
\be
\Gamma_{\alpha\mu\beta}
~=~\Gamma^F_{\alpha\mu\beta}~+~\Gamma^P_{\alpha\mu\beta}
\ee
with 
\be
\Gamma^F_{\alpha\mu\beta}
~=~-(2k+q)_\mu g_{\alpha\beta}
  + 2q_\alpha g_{\mu\beta} -2q_\beta g_{\alpha\mu}
\ee
\be
\Gamma^P_{\alpha\mu\beta}
~=~k_\alpha g_{\mu\beta} + (k+q)_\beta g_{\alpha\mu}~~~.
\ee
 From the simple Ward identity, 
\be 
 \gamma_\alpha k^\alpha
   ~=~ (\not\!p~+~\not\!k)
        ~-~\not\!p
\ee
and using the {\sl equation of motion} for 
the on-shell quarks ($\not\!{p}\psi(p)=0$), 
we note that the contribution from the $\Gamma^P$ 
looks like fig.3(b).
Then the decomposition of the vertex $\Gamma$ into 
$\Gamma^F$ and $\Gamma^P$ 
corresponds to that of the contribution fig.3(a) 
into the two type parts fig.3(b) and fig.3(c).
We redefine the modified two-point function by 
adding the pinch parts from the vertex contributions  
fig.3(b) to the conventional two-point function.
We get the new gauge-invariant self-energy at the 
one-loop level by, 
\be
\hat{\Pi}(q)_{\mu\nu}
  ~=~\left[g_{\mu\nu}
      -{q_{\mu}q_{\nu}\over q^2}\right]\hat{\Pi}(q)
\ee
with
\be
\hat\Pi(q)~=~-bg^2q^2\ln(-q^2/\mu^2)~~~~.
\ee
Here $b$ is the coefficient of ${\cal O}(g^3)$ in 
$\beta(g)$-function,
\be
\beta(g)~=~-bg^3-\sum_{N=1}^{\infty}
b_Ng^{2N+3}
\ee
and $b=11N/48\pi^2$ for SU(N) Yang-Mills theory.

We can construct the gauge independent n-point 
Green's functions by following just the same procedure.

\section{One-loop correction to the $\gamma W W$ vertex}

In ref.\cite{YEHU} E.Yehudai pointed out the 
advantages of the process $e^-\gamma\rightarrow{\nu}W^-$ 
in measuring $W\gamma$ couplings 
in the context of the effective Lagrangian.
In this section, we applied the pinch technique to the process 
$e^-\gamma\rightarrow{\nu}W^-$ to construct the 
gauge-invariant $WW\gamma$ vertex. 
Here we omit the details of the calculation and 
we only show Feynman graphs contributing to 
$WW\gamma$ vertex. 
Diagrams fig.4(a) and fig.4(b) are the tree 
level contributions, 
fig.4(c) and fig.4(d) are the conventional self-energies  
and vertex corrections.
Fig.4(e) is the pinch part from the $e^-\nu{W^-}$ 
vertex and fig.4(f) is the contribution from the box diagram 
to the $WW\gamma$ vertex. 

We will show 
the results of the numerical computations below. 
We put the Higgs mass $M_H=100$ GeV and 
Top mass $M_t=174$ GeV.
In fig.5, we show the dependence on the beam 
polarization, with the beam energy being $\sqrt{S}=300$ GeV.
The cross section for the process 
$e_L^-\gamma_{(\pm)}\rightarrow\nu{W^-}$ are plotted in fig.5(a).
The indices $(\pm)$ on $\gamma$ mean the polarizations 
of $\gamma$ beams.
We plot the ratios of the tree level cross section 
and the cross section including the modified $WW\gamma$ vertex  
in fig.5(b). 
We define the ratio $R$ by,
\be
R~=~{\sigma_{PT}-\sigma_{tree} \over \sigma_{tree}}
\times 100
\ee
where $\sigma_{PT}$ is the cross section which include 
the vertex correction using pinch technique and 
$\sigma_{tree}$ is the cross section at the tree level.
We note that the size of the vertex correction 
is very sensitive to the beam polarizations. 
We plot the energy dependence 
of the unpolarized cross section 
in fig.6(a), and ratios $R$ in fig.6(b) 
at $\sqrt{s}=300$ and $500$ GeV.
In fig.7, we plot  
(a)$\Delta\kappa_\gamma\equiv\kappa_\gamma-1$ 
and (b)$\lambda_\gamma$ 
for the unpolarized case at $\sqrt{s}=300$ and $500$ GeV. 
Since we have calculated the one-loop effects in the 
context of the standard model, 
these quantities depend on $\theta~(t)$. 

\section{Summary}

The pinch technique is an algorithm to construct 
the gauge invariant off-shell Green's function 
in the perturbation theory. 
Using the pinch technique, we can estimate the contribution 
of the one-loop correction to $WW\gamma$ and $WWZ$ vertices  
with the gauge invariance being kept.

We applied the pinch technique to the 
process $e^-\gamma\rightarrow\nu{W^-}$, and 
estimated the radiative corrections to the $WW\gamma$
vertex.  
It is noted that the vertex corrections are sensitive 
to the beam polarization. One-loop contribution to 
$\Delta\kappa_\gamma$ is ${\cal O}(10^{-2})$ 
and $\lambda_\gamma$ is ${\cal O}(10^{-3})$ 
at $\sqrt{s}~=~300~\sim~500$ GeV $e^-\gamma$ collider.

\vspace{20pt}
{\large{\bf Acknowledgment}}\\
One of the author(Y.Y.) would like to thank Prof. K. Sasaki, 
Prof. T. Uematsu and Prof. S. Matsuda for discussions.
This work is supported in part by the Monbusho 
Grant-in-Aid for Scientific Research No. C-05640351 
and No. 050076.


\end{document}